# October 20th, 1993

# Bacteria are not Lamarckian

Antoine Danchin
Institut Pasteur
Unité de Régulation de l'Expression Génétique
Département de Biochimie et Génétique Moléculaire

```
28 rue du Dr Roux - 75724 PARIS CEDEX 15
         Tel 33 (1) 45 68 84 42
         Fax 33 (1) 45 68 89 48
      E-Mail adanchin@pasteur.fr
```

**Instructive influence of environment on heredity has been a debated topic for centuries. Darwin's identification of natural selection coupled to chance variation as the driving force for evolution, against a formal interpretation proposed by Lamarck, convinced most scientists that environment does not specifically instruct evolution in an *oriented* direction. This is true for multicellular organisms. In contrast, bacteria were long thought of as prone to receive oriented influences from their environment, although much was in favour of the Darwinian route (1). In this context Cairns et al. raised a passionate debate by suggesting that bacteria generate mutations oriented by the environmental conditions (2). Several independent pieces of work subsequently demonstrated that mutations overcoming specific defects arised as a consequence of cultivation on specific media (3-7). Two diametrically opposed interpretations were proposed to explain these observations : either induction of mutations instructed by the environment (e.g. by a process involving a putative reverse transcription) or selection of variants among a large set of mutant bacteria generated when stress conditions are present. The experiments presented below indicate that the Darwinian paradigm is the most plausible.**

In order to separate between the two opposite interpretations (instruction vs selection) it was necessary to find as a starting mutation, chosen for its reversion potential, a mutation (i) that was unable to generate true revertants (e. g. a deletion rendering impossible the restoration of original activity) and (ii) that could manifest similar growth properties on *different* growth media, so that influence of the growth medium could be assayed for (a) induction of outgrowing mutants and (b) induction of medium-oriented mutations. Inactivation of the adenylyl cyclase gene of *Escherichia coli* fulfils such requirements. For this reason I used the deletion mutant *cyaΔ854* (8) that renders bacteria unable to ferment a variety of carbon sources. In particular *cya* deficient mutants are unable to ferment maltose, lactose, melibiose, glycerol or mannitol, even when plated on rich media. As a consequence a *cya* mutant yields white colonies on McConkey plates supplemented with these carbon sources (but, as a control, form red colonies on such plates supplemented with glucose). I therefore investigated the fate of colonies obtained on McConkey plates supplemented with glycerol, melibiose, rhamnose, lactose, maltose or mannitol, after one day at 37°C followed by several days at room temperature (18-22°C).

As shown in Table 1 after 8-10 days red papillae started to appear on individual colonies from several plates. Strain TP2000 (*argH1 cyaΔ854 lacΔX74*) was used in the first set of experiments. It carries a lactose deletion that was used as an internal control : no red papillae appeared after one month on McConkey plates supplemented with lactose. The overall number of papillae still grew steadily after one month on all media except lactose and glycerol. The case of glycerol plates is particular, because on such plates growth was found to continue at the low temperature, the colonies slowly increasing in size and displaying a rose, not a white colour. Individual papillae were purified on the original medium at low temperature, then analysed for their fermentation capacity on the other media. Most papillae isolated from maltose, melibiose or rhamnose plates were red on all three media, at low temperature as well as at 37°C (except as expected for melibiose, because of its natural thermosensitivity). They failed to ferment mannitol. In sharp contrast the mannitol papillae were found to be red only on mannitol plates. As already stated no papillae were isolated on glycerol, indicating that papillae could only develop on non growing colonies, but the papillae isolated on maltose, melibiose or rhamnose were found to be red on glycerol at 37°C demonstrating additional capacity for fermenting various carbon sources.

The behaviour of the mutant papillae (which were stable, as shown by repeated streaking on similar plates) was reminiscent of the behaviour of the *crp\** allele of the *crp* gene, coding for a mutated receptor of cAMP that is able to activate transcription of catabolite sensitive operon in the absence of the mediator (9). However, because the original *crp\** mutant had been obtained only from heavily mutated bacteria (9, and A. Ullmann personal communication) the genetic background of the original strain was tested by repeating the experiment in a different background (strain TP7860 : *arg his cyaΔ854 λ*). It yielded similar results. Several P1 lysates were performed on clones isolated from independent plates, and were used to transduce the *crp* region of strain TP2339 (*argH1 cyaΔ854 crpΔ39 lacΔX74*), selecting for Mal$^+$ recombinants. All transductants were found to display the pleiotropic carbon positive phenotype, thus substantiating that the mutation is of the *crp\** type. In one experiment recovery of the wild type *aroB* marker (which is located near the *crp* gene) in a *cya aroB* background was used as a selection for P1 transductants : 38% of the transductants were found to have also recovered a pleiotropic carbon positive phenotype. Subsequently it was found that several different mutants of the *cya* gene (including a Mu insertion) yielded after some time on plates or in stabs bacteria that were able to ferment most if not all carbon sources that depended on the presence of an active CAP-cAMP complex for expression. It was also found that a *recA* derivative of strain TP2000 generated red papillae on maltose, after a lag that was significantly longer (two weeks), demonstrating that *recA* is probably not involved in the generation of papillae, and suggesting that the lag depends on the growth rate of the bacteria : only those colonies that have completely stopped growing seem to be able to generate papillae.

Taken together these results indicate that on most such carbon sources a *crp\** allele can be obtained from papillae outgrowing dormant colonies from *cya* mutants. If the instructive paradigm was to be verified one would have expected to find mostly mutations located in the transcription control regions of the operons specific to each carbon source added to the plates. Indeed, in a few cases (especially on rhamnose plates), mutant types could be obtained, showing specificity towards the original carbon source added to the plates. The *crp\** result cannot be taken as a demonstration of an induction of specific mutations in a carbon source operon, unless one accepts that bacteria possess a magic foresight permitting them to identify this general control gene as the gene that will permit them to escape starvation.

The present experiments rule out the lamarkian interpretation of Cairns' experiments, and substantiate the recent theoretical analysis of Lenski and Mittler (9), but they do not give an explanation at the molecular level of the phenomenon that permits generation of papillae. In this respect it is worth noticing (as already remarked by Cairns or Hall) that papillae grew on colonies only after a certain delay (in a somewhat unpredictable fashion, generally more than five days in our case). In addition it seems important that only those colonies that stopped growing were able to generate papillae (see the glycerol plates experiments). In the same way, when using bacteria that harboured a plasmid permitting synthesis of a very small amount of cAMP (not sufficient to permit the colonies to be red or even rose), no papillae were found. Finally, there was an effect of the nature of the carbon source : on mannitol plates a large number of papillae were obtained, but they did not correspond to *crp\** mutants (no red colonies on the other carbon sources), but to some other mutation permitting expression of the *mtl* operon. All this may suggest that, when in the stationary phase of growth for a significant period of time bacteria are able to enhance the overall mutagenic capacity of a subpopulation, permitting them to generate mutants, some of which are selected as capable of outgrowing the dormant (and probably dying) colony. This should be a further incentive to study gene expression in the late period of the stationary phase.

Acknowledgements

I wish to thank Agnès Ullmann for her interest, and J. Mazé for her help in some experiments.

Table I

Papillae on 1000 colonies after 15 days of strain TP2000 (*cya*-Δ854)

| | Number of papillae | Pleiotropic sugar positive |
|---|---|---|
| Maltose | 32 | + |
| Melibiose | | + |
| Rhamnose | | + |
| Glycerol | no papillae, rose colonies | — |
| Mannitol | >1000 | — |
| Lactose | no papillae, white colonies | — |